\documentclass[useAMS,usenatbib]{mnras}

\usepackage{graphicx} 
\usepackage{times}
\usepackage{epsfig}
\usepackage{epstopdf}
\usepackage{amsfonts}
\usepackage{amsmath}
\usepackage{amsbsy}
\usepackage{bm}
\usepackage{url}
\usepackage{microtype}
\usepackage{rotating}
\usepackage{sidecap}
\usepackage{longtable}
\usepackage{lscape}
\usepackage{rotating}
\usepackage[export]{adjustbox}
\usepackage{amssymb,xcolor}
\usepackage{colortbl}
\usepackage{booktabs}
\usepackage{url}
\usepackage[flushleft]{threeparttable}
\def\apj{ApJ}
\def\apjs{ApJ}
\def\mnras{MNRAS}
\def\aap{A\&A}
\def\apjl{ApJ}

\def\nat{Nature}

\def\araa{Annual Reviews}
\def\apss{Ap\&SS}

\usepackage[T1]{fontenc}
\usepackage{ae,aecompl}

\usepackage{newtxtext,newtxmath}

\title[Sco X-1: Accretion and ultra-relativistic jets]{A connection between accretion states and the formation of ultra-relativistic outflows in a neutron star X-ray binary}

\author[S. Motta \& R. Fender]{Motta, S.E.$^{1}$ \& Fender, R.P.$^{1}$  \\
$^1$University of Oxford, Department of Physics, Astrophysics, Denys Wilkinson Building, Keble Road, OX1 3RH, Oxford, United Kingdom\\
}

\date{Last updated 2017 Apr 4th;}

\pubyear{2015}

\begin{document}
\label{firstpage}
\pagerange{\pageref{firstpage}--\pageref{lastpage}}
\maketitle

\begin{abstract}
The nearby accreting neutron star binary Sco X-1 is the closest example of ongoing relativistic jet production at high Eddington ratios. Previous radio studies have revealed that alongside mildly relativistic, radio-emitting ejecta, there is at times a much faster transfer of energy from the region of the accretion flow along the jet. The nature of this ultrarelativistic flow remains unclear and while there is some evidence for a similar phenomenon in other systems which might contain neutron stars, it has never been observed in a confirmed black hole system. We have compared these previous radio observations with a new analysis of simultaneous X-ray observations which were performed with the RXTE mission. We find that the ejection of the ultra-relativistic flow seems to be associated with the simultaneous appearance of two particular types of quasi-periodic oscillations in the X-ray power spectrum. In contrast, the mildly relativistic, radio-emitting outflows may be associated with flat-topped broad band noise in the X-ray power spectrum.  This is the first time a link, albeit tentative, has been found between these mysterious unseen flows and the accretion flow from which they are launched.
\end{abstract}




\section{Introduction}

Accretion onto compact relativistic objects -- neutron stars and black holes -- is the origin of the highly relativistic jets which we have observed, without fully understanding, for a century now. They transport a vast amount of energy from the central regions of very strong gravitational field, to large distances from the accretor, in the form of highly collimated flows with bulk relativistic motions. The phenomena associated with jets are best studied from a range of viewpoints: jets from active galactic nuclei (AGN) can be observed in exquisite detail at the highest angular resolutions, and in some cases are pointed more or less directly at us. In X-ray binaries (XRB), containing stellar mass black holes (BHs) and neutron stars (NSs), we are able to track how the power and type of jet varies with the state (e.g., accretion flow geometry and optical depth, primarily) and rate of accretion. In the nearest persistent relativistic jet source, the neutron star X-ray binary Sco X-1, a phenomenon has been observed which is unlike anything that has been observed in AGN. In this paper we provide the first evidence for a link between this new jet mode and the accretion flow.

Sco X-1 belongs to a class of accreting systems called Z-sources, which are among the most luminous NS accretors in our Galaxy. They are typically (but not exclusively) persistent, low-mass X-ray binary systems, accreting near or above the Eddington luminosity.
In an X-Ray Colour-Colour diagram or in a Hardness-Intensity diagram (HID) they trace a characteristic Z-shaped pattern formed by three branches -- horizontal branch (HB),  normal branch (NB) and flaring branch (FB) -- which correspond to three accretion ``states'' (see, e.g., \citealt{Hasinger1989}, \citealt{VdK1989}, \citealt{Schulz1993}, \citealt{Kuulkers1994}). An hard apex (HA) separates HB and NB, and a soft apex (SA) separates NB and FB. The X-ray Power Density Spectra (PDS) of Z-sources evolve along the Z track and across the transitions, showing different types of Quasi Periodic Oscillations (QPOs).
kHz QPOs are often detected isolated or in pairs (the so-called ``twins'' kHz QPOs, \citealt{VdK1989}) at several hundreds Hz. Low-frequency QPOs (LFQPOs) appear below $\approx$50 Hz and have been divided in three classes, based on the Z-branch where they are most commonly found: horizontal branch oscillations (HBOs), normal branch oscillations (NBOs) and flaring branch oscillations (FBOs, \citealt{VdK1989}). 
HBOs have been associated with the variation of a ``special radius'' in the accretion flow (typically the disc truncation radius, see e.g. \citealt{Stella1998}, \citealt{Ingram2010}, but also see \citealt{vanderKlis2005} for a review of alternative models), and are believed to be the NS equivalent of the type-C QPOs in BH LMXBs (\citealt{Casella2005,Motta2017}).
NBOs are instead thought to be related to the production of relativistic, transient ejections (see \citealt{Migliari2006}, \citealt{Fender2009}, \citealt{Miller-Jones2012}), and are thought to be the NS equivalent of the type-B QPOs in BH LMXBs.\\
\indent All the known Z-sources are radio bright systems, and their radio emission varies as a function of the accretion state (see, e.g., \citealt{Priedhorsky1986}, \citealt{Hjellming1990a}, \citealt{Hjellming1990b}, \citealt{Penninx1988}, \citealt{Spencer2013}), as a result of a disc-jet coupling similar to that observed in BH LMXBs (Fender et al. 2004, Fender et al. 2009). 
\cite{Migliari2006} showed that while a Z source is moving through the HB towards the NB, the radio power increases. During this phase the radio luminosity is to be ascribed to a compact radio jet, which produces an optically-thick radio spectrum. During the transition from HB to NB, a transient radio jet (responsible for radio thin emission) is launched, with a simultaneous decrease in the compact radio jet power. 
Finally, in the FB, the jet activity is quenched, possibly due to very high accretion rates   (see, e.g.,  \citealt{Bradshaw2003}, but also \citealt{Balucinska-Church2010} and \citealt{Church2012}). As a source moves backwards through the FB and towards the NB, the radio flux increases again. Interestingly, \cite{Migliari2006} associated the HB-to-NB state change with the ejection of transient jets in BH transients, occurring at the transition between the so-called \textit{hard-intermediate state} and the \textit{soft-intermediate state.}

The existence of the aforementioned disc-jet coupling in Z-sources has been established  based on observations of a number of systems, but mainly based on works on Sco X-1 (see \citealt{Fomalont2001}, \citealt{Bradshaw2003}). \\
Cyg X-2 has been observed in radio with the European VLBI Network (EVN) in 2013, while simultaneous X-ray observations were carried out with the \textit{Swift} satellite. During such observations, the source was found in the horizontal branch and a mildly relativistic jet was directly imaged \citep{Spencer2013}.
The `hybrid'\footnote{XTE J1701-462
is the first NS LMXB that clearly showed spectral and timing
properties of both Z and atoll sources (Homan et al. 2007).} source XTE J1701-462 was observed during its Z-phase with ATCA in the radio band, and with \textit{RXTE} in the X-ray band \citep{Fender2007}. Despite the limited sampling, the coupling of radio emission to X-ray states was consistent with that seen in other Z-sources (see \citealt{Migliari2006}).
GX 17+1  was observed simultaneously with the VLA in radio and with RXTE in the X-rays in 2002. \cite{Migliari2007} reported the  evidence of the formation of a radio jet associated with the flaring branch to normal branch X-ray state
(backwards) transition. These authors also found that the radio flux density of the newly formed jet stabilized when the NBO stabilized to its characteristic frequency. This suggested a possible relation between the X-ray variability associated with the NBO and jet formation.
Radio jets have also been seen in the peculiar source Cir X-1, a high-mass NS X-ray binary that has shown erratically a Z-like behaviour (\citealt{Soleri2009}). ATCA observations allowed to resolve an asymmetric arcsecond radio jet, aligned with the larger, more symmetric arcminute-scale collimated structures in the synchrotron nebula surrounding the source \citep{Fender1998}.

Sco X-1 has been observed extensively over the years at all wavelengths and it is among the few NS LMXB where extended radio jets have been unambiguously spatially resolved at sub-milliarcsecond scales. 
Dense radio observations collected in 1999, enhanced by sparser observations obtained earlier, allowed \cite{Fomalont2001,Fomalont2001a} - hereafter F2001a,b - to probe the radio structure of this source, which is formed by a point-like radio core and by radio lobes, ejected in pairs from the system. 
F2001a,b also reported on the presence of a burst of energy emitted from the radio core and travelling towards the lobes at ultra-relativistic velocities, responsible for the radio re-brightening of the radio lobes.
\cite{Fender2004b} reported on an ultra-relativistic energy flow in Cir X-1, similar to that seen in Sco X-1, travelling from the core and  brightening two
radio jet lobes in the approaching jet.
\cite{Migliari2005} found evidence of an analogous ultra-relativistic flow in \textit{Chandra} X-ray observations of the jet source SS433. Two resolved knots in the east jet were seen becoming brighter one after the other, suggesting that a common phenomenon (e.g., a shock wave propagating within the jet itself) might be at the origin of the sequential reheating of the knots. Notably, both in Cir X-1 and SS433 the jets structures are significantly more extended than in the case of Sco X-1 (arcseconds in SS433 and Cir X-1, as opposed to the milli-arcsecods scale structures in Sco X-1).

In this work we investigate the radio and X-ray fast-variability behaviour of Sco X-1, with the aim to explore the coupling between the properties of the accretion flow and the different kind of outflows observed in Sco X-1.
We considered a large amount of data collected by \textit{RXTE} during the radio monitoring carried out in 1999, together with the results by F2001a,b. 
While our radio analysis is necessarily not significantly more advanced than that reported by F2001a,b, the X-ray analysis described in this work has never been performed before, and the results we obtained allow us to improve the understanding of the disc-jet coupling in Sco X-1.

\section{Observations and data Analysis}

The data analysed in this paper were obtained almost 20 years ago. We do not repeat the reduction of the radio data, which is described in great detail, accompanied by many figures and technical information, in F2001a,b. However, it is important to summarize those radio results and the inferred physics, as they are very atypical; this is done in Sec. \ref{ref:whysco} below. We furthermore performed some re-analysis of the radio proper motions in Sec. \ref{sec:radio} and much more extensive analysis of the X-ray data in Sec. \ref{sec:xray}.

\subsection{Summary of the radio properties of Sco X-1}\label{ref:whysco}

Sco X--1 is the brightest X-ray source in the sky and is a well-known, nearby (2.8 $\pm$0.3 kpc, \citealt{Bradshaw1999} and \citealt{Brown2018}) Z-source, constantly accreting close or above the Eddington rate (\citealt{Hasinger1989}, \citealt{Hjellming1990a}, \citealt{Hjellming1990b}).
Sco X-1 is the best-monitored NS LMXB in the radio band. Eight observations with the VLBA, between 1995 August and 1998 August, were planned to optimize the determination of the source trigonometric parallax. Of these, the only observations on consecutive days occurred on 1998 February 27 and 28, from two 6 hours-long VLBA observations, separated by 18 hours.
In 1999 June Sco X-1 was monitored for 56 hours, in a series of seven consecutive 8 hours-long observations amongst three different VLBI arrays: the Very Large Array (VLBA+VLA), the Asia-Pacific Telescope (APT), and the European VLBI Network (EVN). This monitoring constitutes the best and most extensive high angular resolution radio dataset for Sco X-1 obtained to date.
The main findings of F2001a,b relevant for this work are described below. We also summarize such findings in Fig. \ref{fig:schem} in schematic form, and in Fig. \ref{fig:fomalont} in detail. Figure  \ref{fig:schem}, in particular, shows both the observed (right panel) and inferred (left panel) behaviour of Sco X-1, as described in F2001a,b.

\subsubsection{Radio structure} 
F2001a,b showed that the radio imaging of Sco X-1 typically reveals three components: a radio core (blue dot in Fig. \ref{fig:schem}, right panel), corresponding to the binary system, a north-east (NE), and a weaker south-west (SW) lobe (detected less than half of the time), corresponding to the approaching and receding lobe, respectively (labelled as SW and NE lobes in Fig. \ref{fig:schem}).  Both the NE and SW lobes are ejecta launched from the core that move away from it. F2001a,b derived lobe velocities between 0.3 and 0.7c, and core-lobe angular separations between a few and 60\, milli-arcseconds (mas).
Two pairs of lobes were detected in the 1999 radio data, while only one pair was resolved in the 1998 observations. Each pair of moving components persisted up to about 2 days and the generation of pairs occurred often enough that an isolated core component was not commonly observed. The radio lobes emerge from the core and move radially away from it at average apparent speed of (0.45 $\pm$ 0.03)c. The speed of each lobe pair remained unchanged for many hours, although the velocity of different lobe pairs ranged between 0.31c and 0.57c. The properties of the NE and SW lobes strongly suggested that they are intrinsically similar, but they differ in terms of relativistic aberration. This allowed F2001a,b to constrain the inclination of the jets with respect to the line of sight to 44$\pm$7 degrees, under the assumption that the lobes are launched symmetrically with respect to the binary \citep{Blandford1977}. This estimate was done taking the weighted average velocities of the approaching and receding lobes.
 
\subsubsection{Flaring activity} 
Unusually, although not uniquely, as well as core radio flaring, flaring (rebrightening) behaviour is also observed in the moving lobes of Sco X-1. 
The radio data can be seen in Fig. \ref{fig:fomalont}, adapted  from \cite{Fomalont2001}, but with the addition of our re-analysis of the radio data, which will be discussed in Sec. 2.2. As the figure shows, in the 1999 data the radio core showed 4 major flares (C1, C2, C3, C4 in Fig. \ref{fig:fomalont}, bottom panel). The first of the two NE lobes resolved in 1999 (NE lobe \#1) showed one flare (N1), while the second NE lobe (NE lobe \#2) showed three flares (N2, N3, N4, in Fig. \ref{fig:fomalont}, bottom panel). Only the second of the two SW lobes (SW lobe \#2) showed a clear flare (S3), while the first SW lobe (SW lobe \#1) was only barely detected.

\subsubsection{Ultra Relativistic Outflows} 
\begin{figure*}
\centering
\includegraphics[width=1.0\textwidth]{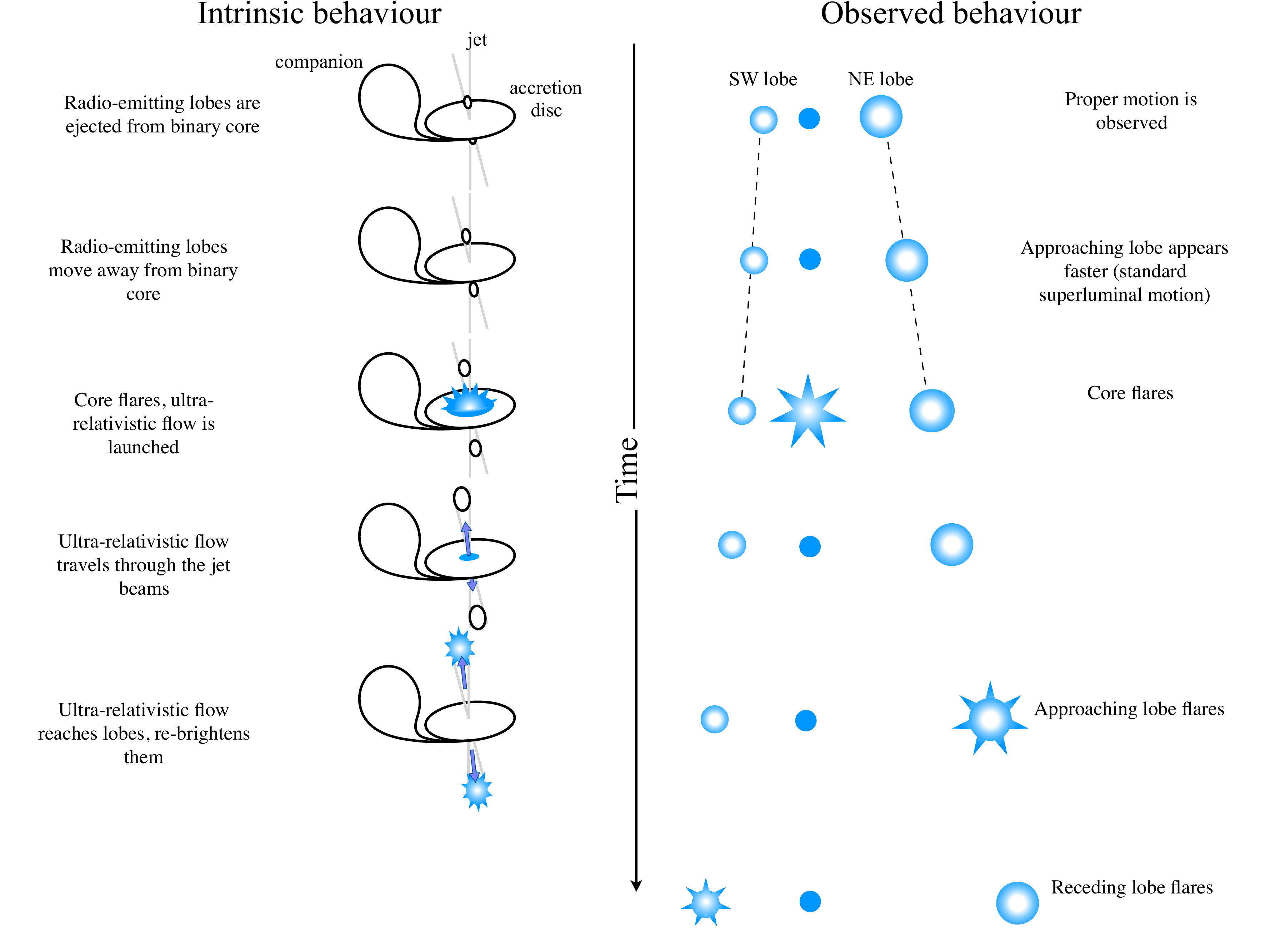}
\caption{Schematic representation of the observed and inferred core, lobe and beam behaviour of Sco X-1. The left panel illustrate the events in the binary framework, while the right panel shows what the observers sees. 
}
\label{fig:schem} 
\end{figure*}

\begin{figure*}
\centering
\includegraphics[width=1.0\textwidth]{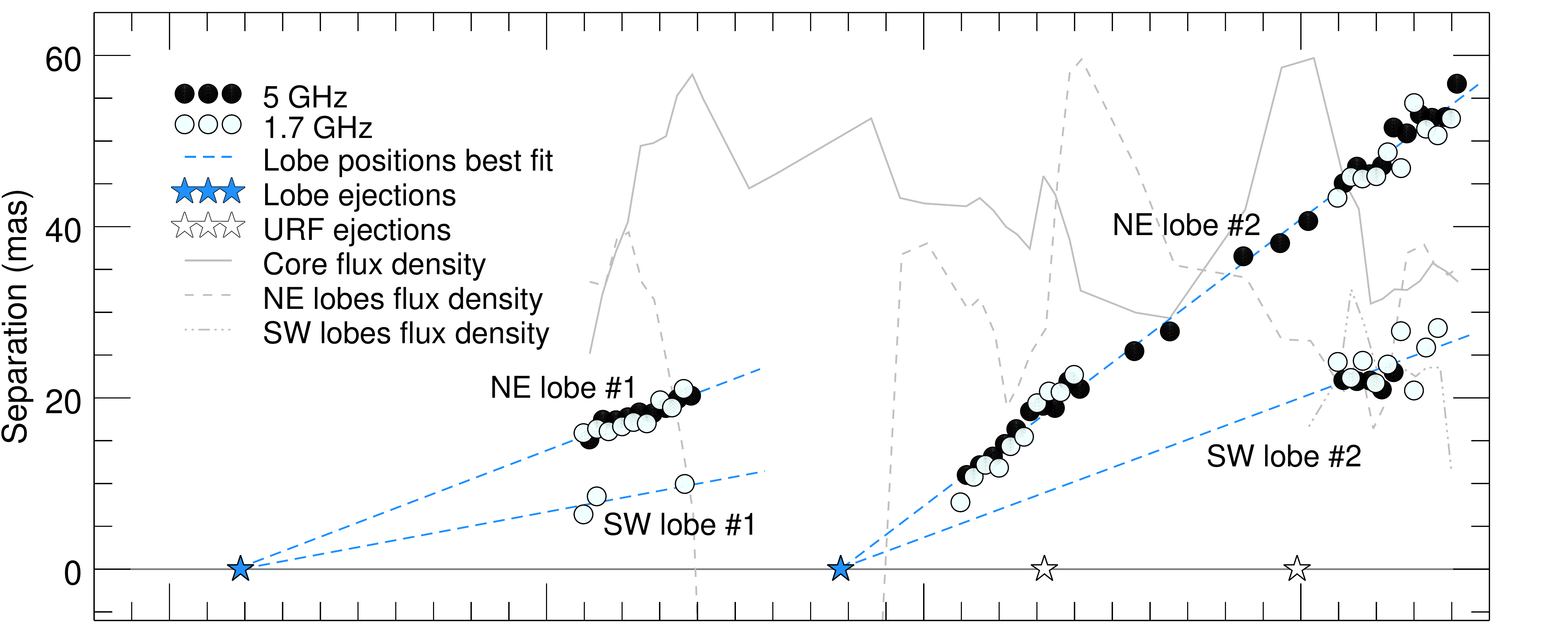}
\includegraphics[width=1.0\textwidth]{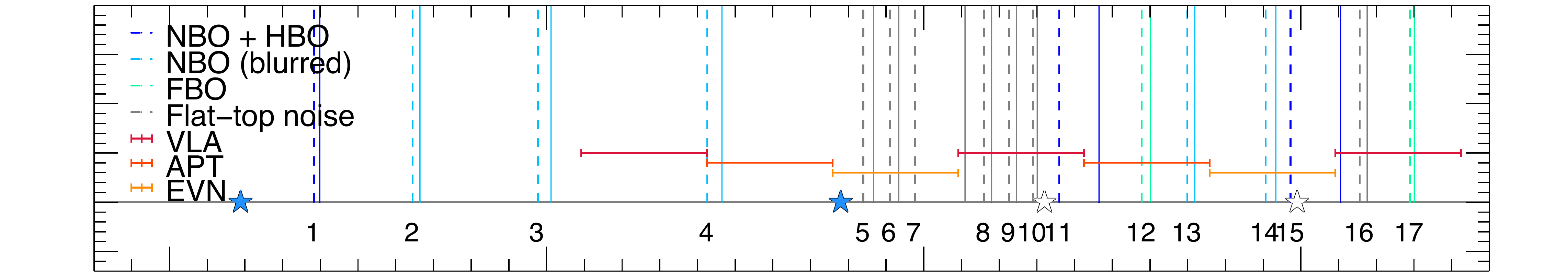}
\includegraphics[width=1.0\textwidth]{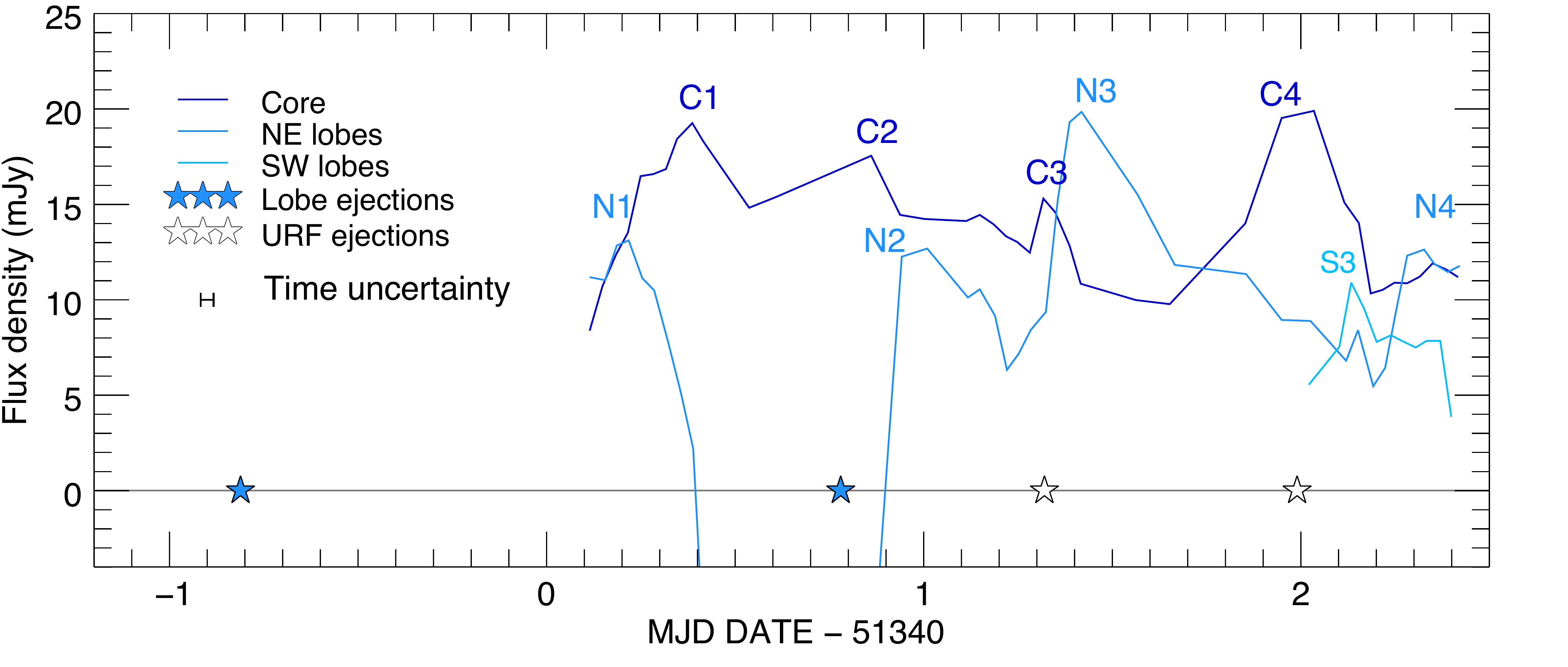}
\caption{\textbf{Top panel:}  Time evolution of the separation of the SW and NE lobes from Sco X-1 during observations in 1999. The separation is expressed in milli-arcseconds (equivalent to $4 \times 10^8$ km for the case of Sco X-1). Data points mark the relative position of the ballistic ejecta as a function of time at 5 GHz (black) and 1.7 GHz (white). The dashed blue lines correspond to our fits to the Core-North-East and Core-South-West lobes separation as a function of time, based on the data in F2001a,b, which intercept separation = 0 at the ejection time. The grey solid, dashed and dot-dashed lines indicate the core, NE lobe and SW lobe flux density (rescaled) shown in the bottom panel. The blue and white stars mark the ejection of the lobes and of the ultra relativistic flow, respectively. Note that the four star symbols are larger than the uncertainties associated to the ejection time (0.02 days, see text for details). This figure has been adapted from Fomalont et al. 2001b (see their Fig. 3, which we digitized).
\textbf{Middle panel:} X-ray and radio observations. The vertical lines indicate the start and stop times (thick dashed lines and thin solid lines, respectively) of the \textit{RXTE} observations considered in this work. The colour of the lines indicate the main QPO type observed in the X-ray PDS (see legend for a classification). The number below each vertical dashed line corresponds to the RXTE observation number indicated in Tab. \ref{tab:log} and to a panel in Fig. \ref{fig:PDS}, where we report all the PDS extracted during the period covered by the radio observations. The horizontal lines indicate the exposure times of the radio observations considered here, divided by instrument.
\textbf{Bottom panel:} flux density of the components of Sco X-1 during the radio observations reported by F2001a,b. The two lobes ejections and the two invisible ejections are indicated as in the top panel. The extent of the average time-uncertainty on the radio events is show in the legend. This corresponds to the approximate exposure of the snapshots used to measure the flux of the radio components of Sco X-1 in F2001a,b.
}
\label{fig:fomalont} 
\end{figure*}

F2001a,b inferred the presence of the ejection of an invisible (i.e., not directly imaged in the radio or X-ray data), \textit{ultra-relativistic flow} (URF) from the radio core of the system to the radio lobes. They showed that the flux density variations of the radio core are correlated with the flux density variations of the lobes assuming that an energy flow moves from the core to the lobes through a symmetric twin beam connecting them, with a speed greater than $\approx$0.95\ c. F2001a,b suggest that as the radio core flares due to an accretion event occurring in the binary system, a burst of energy associated with this event travels down the twin-beam connecting core and lobes at a highly relativistic velocity. This increased energy flux intercepts the NE and SW lobes inducing a flare in both components (see Fig. \ref{fig:schem}, left panel). Because of the angle to the line of sight, the approaching NW lobe always re-brightens before the receding SW lobe (see Fig.\ref{fig:schem}, right panel). At least two URF ejections - separated by $\approx$14 hours - are inferred to occur during in the 1999 radio observations.

\subsection{Partial reanalysis of radio proper motions}\label{sec:radio}

Our main aim was to perform extensive comparison between the radio and the X-ray properties of Sco X-1, therefore we considered the results from the radio observations performed in 1999 (F2001a,b), which were carried out simultaneously with extensive X-ray monitoring with RXTE (see Sec. \ref{sec:xray}). Previous observations did not always enable to resolve the radio components of Sco X-1 and to probe their fast flux-density variability due to limited sensitivity and spatial resolution.

The VLA/EVN/APT data collected in 1999 cannot currently be re-analysed from scratch, and the results from the analysis by Fomalont and collaborators are not electronically available. Therefore, we digitized\footnote{We recovered the values obtained by \cite{Fomalont2001a} by measuring the x,y coordinate of each point in their Fig. 7 (top panel), and Fig. 16 (top panel). We based our re-analysis of the radio data on such recovered values.} the data reported in \cite{Fomalont2001a} using their Fig. 7 (top panel) and Fig. 16 (top panel). While this still allowed us to directly compare the radio and X-ray data, the digitization obviously involves the unavoidable introduction of systematics that limits in-depth analysis of the radio data. In order to take this into account, we adopted conservative time uncertainties on the radio measurements (i.e. 0.02 MJD on each radio detection, see below for further details).

We inferred the ejection times of the two pairs of lobes detected in the 1999 data (SW and NE lobes \#1 and \#2)  by fitting the separation of the moving components with respect to the radio core versus time, then extrapolating the time when the lobes emerged from the radio core.
Since the SW lobes are most of the time too faint to be clearly resolved, in our fits we only considered the positions of the NE lobes to extrapolate the ejection time of the lobes. 

F2001a,b reported a change in speed of the NE lobe \#2 at an MJD of about 51341.4, which translates in a break to the power law fit used to model the data. We note, however, that based on the values we recovered from \cite{Fomalont2001a}, such a velocity change does not appear to be highly significant: from visual inspection of the data we note that the set of positions of the NE lobe \#2 might be equally well described by a linear fit. 
Furthermore, the positions of the NE lobe \#2 collected at 1.7 GHz and at 5 GHz do no completely agree, especially at short core-lobe separations. 
Therefore, we decided to adopt a strategy different from that followed by F2001a,b. We initially considered only the positions of the  NE lobe \#2 obtained from the observations taken at 5 GHz data, which are expected to be affected by smaller uncertainties than those from the 1.7 GHz data. In the other three cases (NE  lobe \#1 and SW  \#1 and \#2) we considered both the positions at 1.7 and 5 GHz.  We fitted to the positions a linear relation in the form $y = b t$, where $y$ is the lobe-core distance and $t$ is the time at which the position of a given lobe is taken (see Fig. \ref{fig:fomalont}, top panel).
In order to better constrain the fits to the SW  \#1 and \#2 positions (both fainter and detected for shorter time intervals), we assumed that the launch time of the NE and the SW lobe is the same, and that the two lobes are launched symmetrically with respect to the binary. Based on these fits, we solved the jet proper motion equations (see, e.g., \citealt{Fender2003b}), obtaining the proper velocities of the lobes, as well as their ejection  angle with respect to the line of sight. 

In order to compare our results to those by F2001a,b, we repeated the fit to the NE lobe \#2 positions, this time considering the positions obtained at both 1.3 and 5 GHz, but only collected before 51341.4, i.e. preceding the apparent change in the lobe velocity reported by F2001a,b, in order to infer the launch velocity. We find that the apparent motion of the NE lobe \#2 can be described equally well by one simple linear fit (fit shown in Fig. \ref{fig:fomalont}), or by a broken linear relation (as originally suggested by F2001a,b, see Fig. \ref{fig:positions}).  
This, however, implies that, depending on which type of fit one chooses, the ejection time of the second pair of lobes varies by a few hours, with the ejection occurring earlier or later for a single or broken linear fit, respectively.

\cite{Fomalont2001a} reported the velocities and ejection times for the URF that caused the flares in the SW and NE lobes correlated with the core flares. 
We re-estimated the velocities of the URFs using equation 13 in \citealt{Fomalont2001a} and using the velocity of the NE lobe \#2 and the ejection angle with respect to the line of sight that we obtained through our analysis. Following F2001a,b, we assumed that the URFs are launched at the peak-time of each core flare correlated with lobe flares (i.e. core flare C3, correlated with lobe flare N3, and core flare C4, correlated with lobe flares N4 and S4). Even though this is not necessarily the most correct  assumption, it allows us to mitigate as much as possible the introduction of uncertainties in the measurement of the time delay between core and lobe flares, required to estimate the URF velocity. 
We associated a time uncertainty of 0.02 days to all the radio measurements, corresponding approximately to half of the duration of the radio snapshots used in F2001a,b, which lasted between $\approx$45 min and $\approx$1 hr. In our analysis we did not directly consider the uncertainty on the positions of the radio lobes (which could not be extracted accurately through data digitization), as the time uncertainties dominate over the spatial ones.

\begin{figure*}
\centering
\includegraphics[width=0.98\textwidth]{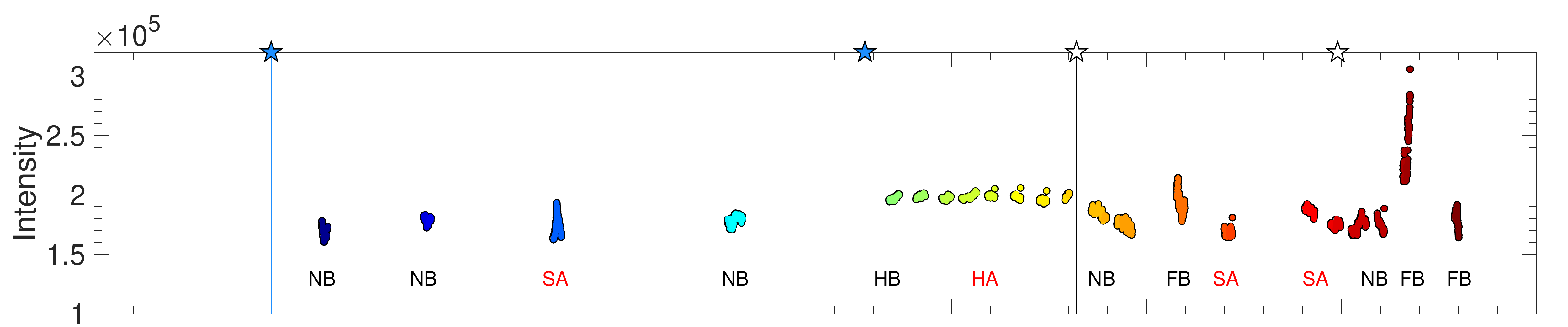}
\includegraphics[width=0.98\textwidth]{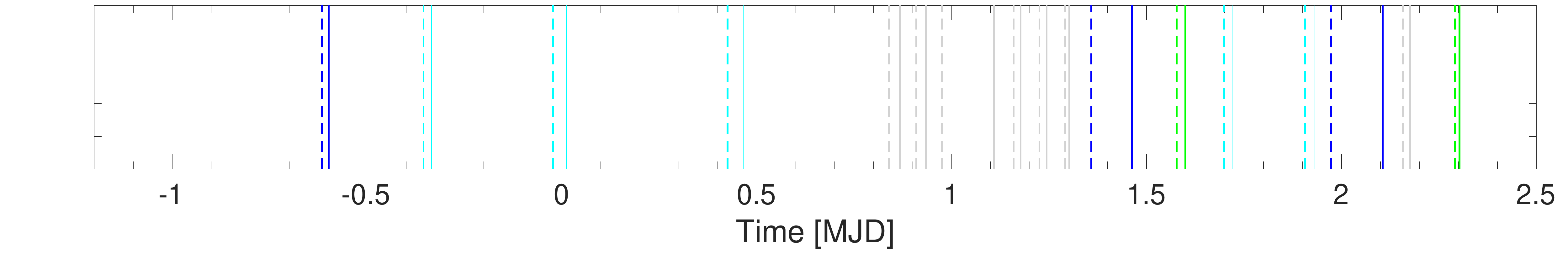}
\caption{Top Panel: X-ray light curve (16s resolution) extracted from the 17 RXTE observations considered in this work. The light curve is colour-coded according to time, in order to facilitate the inspection of the HID in Fig. \ref{fig:CCD}. The times of the ejections are marked with dashed lines and with stars on the plot, following the same criteria used in Fig. \ref{fig:fomalont}. Bottom panel: start and end time of the X-ray observations. Lines are colour coded according to the main QPO(s) observed in the PDS, as in Fig. \ref{fig:fomalont} (middle panel).
}
\label{fig:LICU} 
\end{figure*}

\begin{figure}
\centering
\includegraphics[width=0.48\textwidth]{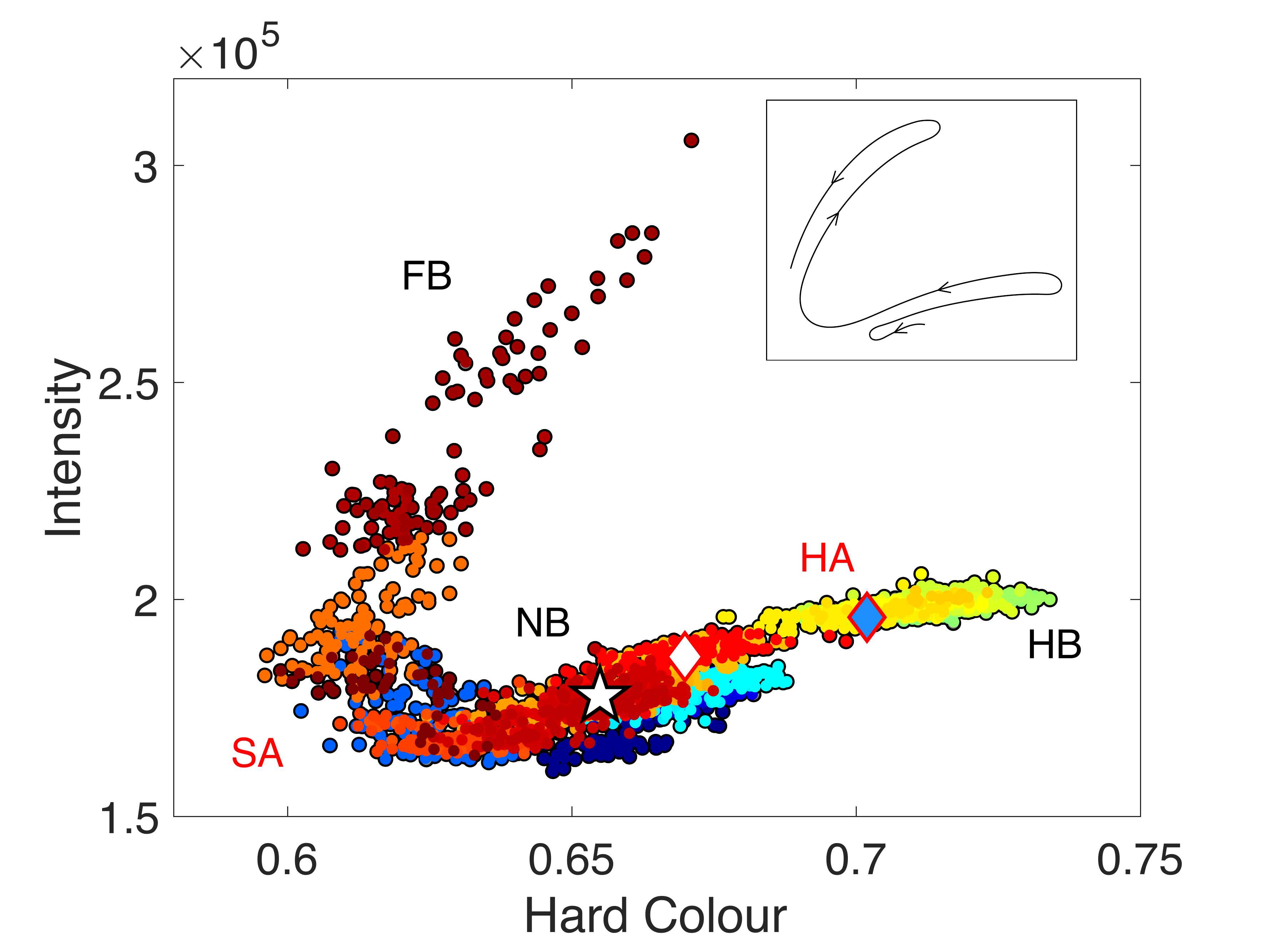}
\caption{Hardness-Intensity diagram of Sco X-1 extracted form the 17 RXTE Observations considered in this work. Each point in the plot corresponds to a data segment 16s long. Points are colour-coded according to time as in Fig. \ref{fig:LICU}, to facilitate the inspection of the diagram. The three branches of the Z track are indicated on the plot, along with with the location of the transitions between them (HA and SA). The inset shows the source evolution along the Z-track. 
The red-edged blue diamond corresponds to the approximate location of the lobe ejection \#2 (based on RXTE observations \#5 and \#6). The red-edged white diamond corresponds to the approximate location of the ejection of the URF The \#1 (based on RXTE observations \#10 and \#11). The white stars corresponds to the exact location of the ejection of the URF \#2 (based on RXTE Observation \#15). }
\label{fig:CCD} 
\end{figure}

\subsection{X-ray data}\label{sec:xray}

During its 16 years of activity \textit{RXTE} observed Sco X-1 extensively. Observations simultaneous with the radio observing runs were performed on August 3rd, 1997 (MJD 50663, one observation), February 27th and 28th (MJD 50871 and 50872, two observations) and June 11th and 12th 1999 (MJD 51340 to 51342, seventeen observations).
In this work we only considered the \textit{RXTE} observations taken in 1999 (quasi) simultaneously with the radio data (see Tab. \ref{tab:log}), which allowed us to perform a systematic comparison between the results from the X-ray and radio observations. 

We produced a light curve (Fig \ref{fig:LICU}) and an HID (Fig \ref{fig:CCD}) with a 16s resolution from the seventeen RXTE observations collected in 1999, using \textsc{Standard 2} data, characterised by a low time-resolution (16s), but a good energy resolution (129 channels covering the nominal energy range 2-120 keV). We extracted count rates in the energy bands 
A $\approx$8-10 keV, B $\approx$10-19  keV, and C $\approx$2-19 keV (absolute channels 18-23, 24-43, 3-43, respectively), and we defined the hard colour as B/A. The HID was produced by plotting the intensity, i.e. the count rate in band C, versus the hard colour. The above count rates have been extracted directly from the \textsc{Standard 2} event files, thus have not been background corrected. We note, however, that the average \textit{RXTE/PCA} background is of the order of 20 counts/s, while the average count rates form  s Sco X-1 in the above bands are of the order of 10$^4$-10$^5$ counts/s, which makes the background negligible.

For each observation we computed PDS from \textit{RXTE}/PCA data using custom software under \textsc{IDL}\footnote{GHATS, http://www.brera.inaf.it/utenti/belloni/GHATS\_Package/Home.html} in the energy band 2-120 keV (absolute PCA channel 0 to 249), from \textsc{BINNED} (covering the energy range $\approx$2-20 keV) and \textsc{SINGLE BIT} mode (covering the energy range $\approx$20-120keV) data combined. This allowed us to extract PDS covering the entire available energy range.
We used 16s-long intervals and a Nyquist frequency of 2048 Hz to produce PDS normalized according to \cite{Leahy1983}, which we averaged obtaining one PDS per \textit{RXTE} observation (shown in Fig. \ref{fig:PDS}).
We fitted each PDS with the \textsc{XSPEC} fitting package (typically used for X-ray spectral analysis purposes) using a one-to-one energy-frequency conversion and a unity response matrix to allow us to fit a power vs. frequency spectrum as if it was a flux vs. energy spectrum. Following \cite{Belloni2002} we fitted the noise components with a number of broad Lorentzians, and QPOs with one or more narrow Lorentzians. We also added a constant component to take into account the contribution of the Poisson noise.

\section{Results}\label{Sec:radioresults}

\subsection{Radio data analysis results}

In Figure \ref{fig:fomalont} (top panel) we show the separation of the NE and SW radio lobes from the radio core (in mas units) as a function of time, as reported by \cite{Fomalont2001a}, together with the velocities derived through our re-analysis. To facilitate the inspection of the radio behaviour of the source, in the same plot we also show the flux density variability of the core and lobes as a function of time, displayed separately and with flare labels in the bottom panel of Fig. \ref{fig:fomalont}. The average error bar associated to the time of detections is shown in the legend of the bottom panel. Note that the same uncertainties apply to each measurement of the radio lobes flux densities, and also on the lobe positions reported in the top panel of Fig. \ref{fig:fomalont}.

In the top panel of Fig. \ref{fig:fomalont} we also mark the results of the best fit to the positions of the NE lobes, and the result of the extrapolation to obtain the ejection times (blue dashed lines in Fig. \ref{fig:fomalont}). Based on our results, the ejection times of the two pairs of lobes resolved in the radio data occurred on MJD 51339.254 and MJD MJD 51340.779, respectively. These ejections are marked by blue stars in all the panels of Fig. \ref{fig:fomalont}. 
The ejections of the NE and SW lobes \#1 occurred about 20 hours before the beginning of the VLA radio observations (started on MJD 51340). The ejections of the NE and SW lobes \#2, instead, occurred about 7 hours before the first detection of the second pair of lobes (based on a linear fit, see below). The proper motion of the NE and SW lobes are reported in Tab. \ref{tab:fit_par}, as well as their inferred proper velocities and launching angle, as obtained solving the jet proper motion equations. 
The use of a linear fit to the positions of the NE lobe \#2 implies a slightly slower
proper motion for both the NE and SW lobe \#2 with respect to what reported by F2001a,b (we obtained a proper motion of $\approx 33.5$ mas/day, as opposed to $\approx 41.7$ mas/day reported by F2001a,b). Interestingly, we find that both the lobes velocity and launch angle change for different lobe pairs. In Table \ref{tab:fit_par} we report the values inferred through our calculations. For comparison, we also report the values obtained by F2001b based on data collected in 1998 (not considered in this work), where a third pair of lobes have been detected and for which the proper motion could be measured. 

\begin{figure}
\includegraphics[width=0.50\textwidth]{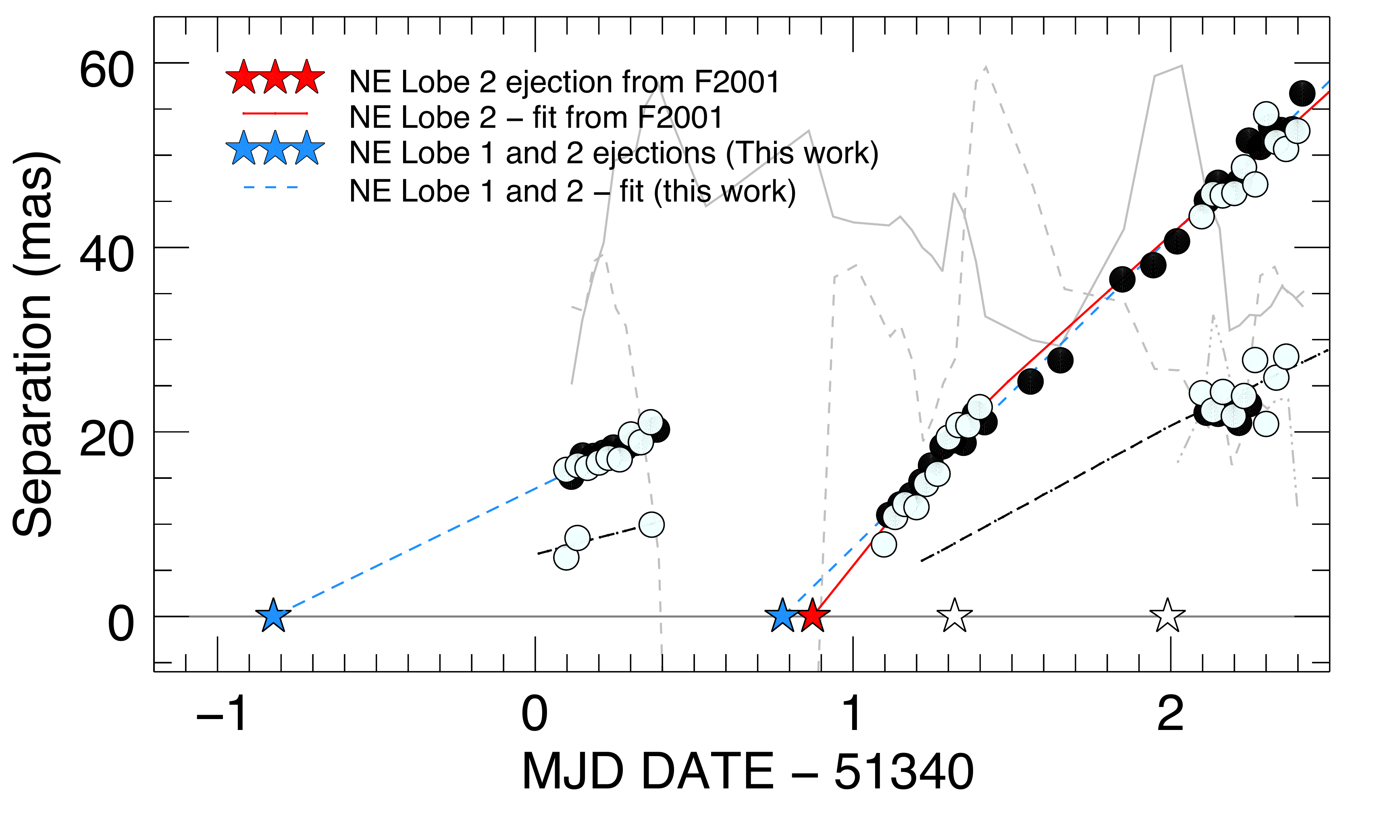}
\caption{Same as in Fig. \ref{fig:fomalont} (top panel), but with the motion of the NE lobe \#2 fitted with a broken linear fit (indicated with a red line) as in F2001a,b. The inferred ejection time (MJD 51340.873, indicated with a red star) follows the ejection time based on a simple  linear fit (second blue star from the left) by a about two hours. The blue dashed line marks the linear fit shown in \ref{fig:fomalont} (top panel) for comparison. }
\label{fig:positions}
\end{figure}

\indent As mentioned in Sec. \ref{sec:radio}, the ejection time of the NE lobe \#2 varies depending on the type of fit adopted to describe the motion of this lobe, i.e. a simple linear fit, or a broken linear fit (following F2001a,b). The ejection time obtained through a broken linear fit (see \ref{fig:positions}) is MJD 51340.873, i.e., about two hours later than the ejection inferred based on our simple linear fit. The latter ejection time is marked with a red filled star in Fig. \ref{fig:positions}, where we also plot the results of the simple linear fit for comparison (note that the portion of the fit to the NE lobe positions after MJD 51341.4 is not show, and the reader is further referred to F2001a,b for details). Note that we do not report in Tab. \ref{tab:fit_par} the results of a broken linear-fit to the positions of the NE lobe \#2, as they are consistent with those reported by F2001a,b. We therefore refer the reader to the values reported in their work.  In both cases, however, given the uncertainties on the times of the radio events, the ejection  of the second lobe pair occurs approximately at the time of core flare C2. 
Adopting the Occam's razor approach, we based our subsequent analysis and discussion on the results of the linear fit to the positions of the NE lobe \#2, i.e. taking the ejection time obtained thorough a linear fit. 

%

%

\begin{table*}																
\caption{Results of the best fit to the projected positions of the approaching (NE) and receding (SW) lobes lobes \#1 and \#2. Note that the proper motion $\mu$ corresponds to the b parameter of the linear fit we performed on the data. We indicate the inferred ejection time of the two pair of lobes, and the proper motion of the four lobes detected, as well as a number of quantities obtained solving the proper motion jet equations: the unique product $\beta cos \theta = (\mu_{app}-\mu_{rec})/(\mu_{app}+\mu_{rec})$, and $\beta$ and $\theta$ obtained assuming symmetry in the lobe ejection.}  
\begin{threeparttable}
\centering		
\begin{tabular}{c c c c c c c}										

\hline		
Event	 					& Ejection time 		&	$\mu_{app}^{\dagger}$  ( b) 	 & $\mu_{rec}$    & $\beta cos \theta$ &	$\beta$   	   & $\theta$	  \\
	 					    & [MJD]         		&	[mas/day]    	 & [mas/day]      &                    &	          	   & [degrees]	          \\

\hline			
Lobe pair \#1 				& 51339.176$\pm$0.005 	&	16.81$\pm$0.06   & 8.23$\pm$0.02  &  0.343$\pm$0.002   &  0.387$\pm$0.009  & 28$\pm$3             \\
Lobe pair \#2		 		& 51340.780$\pm$0.001	&	33.480$\pm$0.001 & 16.32$\pm$0.02 &  0.3445$\pm$0.0004 &  0.49$\pm$0.03   & 46$\pm$3              \\
Lobe pair \#3$^{\ddagger}$	& $\approx$50871		&	26.64$\pm$0.02   & 14.40$\pm$0.08 &  0.298$\pm$0.003   &  0.42$\pm$0.02    & 45$\pm$3             \\

\hline				

\end{tabular}
\begin{tablenotes}
\item $\dagger$ Corresponding to parameter b in the fit to the positions of the approaching and receding lobes.
\item $\ddagger$ We indicate as lobe pair \#3 the pair of approaching and receding lobes detected in the VLBI data collected in 1998 and reported by \cite{Fomalont2001a}, which have not been re-analysed in this work. We thus use the values reported by \cite{Fomalont2001a}. 
\end{tablenotes}
\end{threeparttable}\label{tab:fit_par}
\end{table*}	

%

We estimated the velocity of the URFs by assuming that the URF ejection times correspond to the peak times of flares C3 and C4, and measuring the relative delay between core flare C3 and lobe flares N3 and S3, and between core flare C4 and lobe flare N4. 
By taking the lobe proper velocity and launch angle obtained above, we obtained a velocity $\beta_{\rm URF} \approx 1$ for both the URFs launched, with a strict lower limit (99\% confidence level) of $\beta_{\rm C3-N3}^{low} = 0.81$ and $\beta_{\rm C3-S3}^{low} = 0.88$, from the correlated flares C3--N3 and C3--S3, respectively, and of $\beta_{\rm C4-N4}^{low} = 0.87$ from the correlated flares C4--N4. These values are consistent with those reported by F2001a,b, who derived a weighted lower limit to the velocity of the URF of $\beta_{URF} > 0.95$.
In Tab. \ref{tab:rel_par} we indicate the apparent velocity of the URF as obtained from the three correlated flares, the corresponding proper velocities, and the associated lower limits to the Lorentz factors $\Gamma$. We stress that these are conservative lower limits to the Lorentz factor, under the assumptions of our model, and that there is essentially no upper limit to $\Gamma$.

\begin{table}									
\centering									
\caption{Correlated core-lobe radio flares considered, apparent velocity of the URF, URF intrinsic velocity  ($\beta_{URF}$) and associated URF minimum Lorentz factors ($\Gamma^{min}$).}  
\begin{tabular}{c c c c}													

\hline		
Flares			& Apparent URF velocity       &  $\beta_{URF}$	&	$\Gamma^{min}$    					\\
                & 		 [mas/day]            &   				&	at 3$\sigma$ confidence  			\\

\hline			
C3 - N3  		&  252     					  &	 1.1$\pm$0.1    		&   1.69    		  	\\
C4 - N4  		&  166.9       				  &	 1.03$\pm$0.05    		&  	2.04  			 	\\
C3 - S3  		&  26.9   	  				  &	 1.06$\pm$0.06			&   2.12 		  	  	\\

\hline				

\end{tabular}\label{tab:rel_par}									
\end{table}						

\subsection{X-ray data analysis results}

In Figure \ref{fig:fomalont} (middle panel) we indicate the start and stop time of the \textit{RXTE} observations considered in this work (thick dashed lines and thin solid lines, respectively). The colour of the lines marks different types of QPOs seen in the X-ray PDS: dark blue for simultaneous NBO and HBO, light blue for an NBO, light green for an FBO, grey if only a flat-topped noise component is detected (without QPOs below $\sim$100 Hz). The number below each vertical line corresponds to a row in Tab. \ref{tab:log} and to a panel in Fig. \ref{fig:PDS} in the Appendix, which presents the X-ray power spectrum for each observation. 
For reference, we also indicated the start and stop of the radio observations described in the previous sections with horizontal line (red is VLA, dark orange is APT, orange is EVN).

Figure \ref{fig:LICU} (top panel) and \ref{fig:CCD} show the light curve and the HID, respectively, extracted from the RXTE data. Both figures are colour-coded according to time, to allow the reader to easily follow the evolution of the source along the Z-track in the HID. In the top panel of Fig. \ref{fig:LICU} we also indicate the position of the source along the Z-track based on the HID. The bottom panel of Fig. \ref{fig:LICU} is a replica of the middle panel of Fig. \ref{fig:fomalont}, which we report to facilitate the inspection of the X-ray light curve).
The colour of the lines indicate the main type of QPO observed in the average PDS extracted from each observation. In Fig. \ref{fig:CCD} we label the areas of the HID according to the historical state classification (horizontal, normal or flaring branch, and hard and soft apex).
The position of each observation on the Z-track can be accurately found 
based on the HID. The detection of particular types of QPOs, however, provides additional information on the state of the source. As Fig. \ref{fig:LICU} and \ref{fig:CCD} show, in the X-ray dataset considered here, Sco X-1 samples the entire Z-track in a time interval shorter than a day.

In Fig. \ref{fig:PDS} in Appendix \ref{App:PDS} we show the PDS produced from the 17 RXTE observations considered in this work. The inspection of the dynamic power density spectra confirms that the overall power density distribution does not change significantly during any of the observations considered, despite the fact that some of the observations covered long time intervals (> 2 hr). Therefore, we are showing only one average PDS for every observation, enough for the purposes of this paper, as they show the main QPO seen at the time of each observation\footnote{Note that this approach is not necessarily correct in cases where the power density distribution from a variable source changes significantly on short time scales: in these cases producing a PDS over a time interval significantly longer than said time scale would result in a meaningless PDS. }. We note, however, that some observations (e.g. Obs. 15) will suffer from data mixing, as QPOs that appear at different times, will show simultaneously in the PDS.
In Table \ref{tab:PDS_fits} we report the centroid frequencies of the QPOs detected in the observations considered here. We note that in the case of FBOs, the centroid frequency reported is to be intended as an average frequency, as this type of QPO is know to vary significantly around a central value on time scales generally shorter than the average length of a typical RXTE pointing.

We detected several QPOs, such that the majority of the observations considered shows at least a QPO. In particular, we observed simultaneous NBOs and HBOs in three observations (1, 11 and 15), an NBO that evolves into an FBO in three observations (3, 13 and 17), an isolated NBO in three observations (2,3, 4 and 14), an isolated FBO in one observation (12), and no QPO, but only flat-topped noise in the remaining observations. In this dataset, we did not detect isolated HBOs, despite the fact that they are observed - though not frequently - in the light curve of Sco X-1.

\begin{table}									
\centering									
\caption{Centroid frequency of the QPOs detected in the PDS shown in Fig. \ref{fig:PDS}.}  
\begin{tabular}{c c c c c}													
\hline \hline			
 \#	& \textit{RXTE} Obs.ID &	NBO Freq.	&	HBO freq.	&	FBO freq.		\\
		&					&	[Hz]			&	[Hz]			&		[Hz]	\\
\hline		\hline				
{\bf 1} & 40706-02-01-00 &	6.03$\pm$0.03	&	44.6$_{-0.7}^{+0.6}$	&	-           	\\
{\bf 2} & 0706-02-03-00 &	6.3$\pm$1	&	-		    &	-           					\\
{\bf 3} & 0706-02-06-00 &	9.14$\pm$0.08	&	-           &	16.7$\pm$0.2 	\\
{\bf 4} & 40706-01-01-00 &	8.77$_{-0.5}^{+0.6}$	&	-			&   -			 	\\
{\bf 5} & 40706-02-09-00 &	-			&	-           &	-           	\\
{\bf 6} & 40706-02-08-00 &	-			&	-           &	-           	\\
{\bf 7} & 40706-02-10-00 &	-			&	-           &	-           	\\
{\bf 8} & 40706-02-12-00 &	-			&	-           &	-           	\\
{\bf 9} & 40706-02-13-00 &	-			&	-           &	-           	\\
{\bf 10} & 40706-02-14-00 &	-			&	-           &	-           	\\
{\bf 11} & 40706-01-02-00 &	5.84$_{-0.03}^{+0.02}$	&	44.7$_{-0.8}^{+0.9}$ &	-          \\
{\bf 12} & 40706-02-16-00 &	-           &	-	&	17.2$\pm$0.2           	\\
{\bf 13} & 40706-02-17-00 &	8.9$_{-0.2}^{+0.1}$	&	-           &	12.2$_{-0.3}^{+0.2}$ 	\\
{\bf 14} & 40706-02-18-00 &	8$\pm$1	&	-           &	-           	\\
{\bf 15} & 40706-02-19-00 &	6.18$\pm$0.01	&	43.1$\pm$0.7	&	-           	\\
{\bf 16} & 40706-02-21-00 &	-			&	-           &	-           	\\
{\bf 17} & 40706-02-23-00 &	9.2$_{-0.3}^{+0.4}$	&	-           &	17.0$\pm$0.4 	\\
\hline				

\end{tabular}\label{tab:PDS_fits}									
\end{table}						

\subsection{Comparison of radio and X-ray behaviour}

We summarize in the following the results of the comparison of the X-ray timing analysis and of the radio analysis, focussing on when X-ray observations were made near the inferred ejection times of either lobes or the URF.

\subsubsection{Ejection of lobes \#1}
Observation \#1 started on MJD 51339.384, about 3 hr after the ejection of NE and SW lobes \#1. The corresponding PDS (displayed in Fig. \ref{fig:PDS}, panel 1) shows both a strong NBO and an HBO (at $\approx$ 6 and 44 Hz, respectively). The \textit{RXTE} observation taken immediately after (Obs. \# 2) started on MJD 51339.645. The corresponding PDS still shows an NBO  with centroid frequency at $\approx$6 Hz and a hint of an HBO at about 45 Hz, both at frequencies consistent with those of the QPOs observed in observation \#1 (see Tab. \ref{tab:PDS_fits}). The PDS extracted from observations \#3 and \#4 show an NBO at a slightly higher frequency ($\approx$8-9 Hz, see Tab. \ref{tab:PDS_fits}) with respect to that measured in observations \#1 and \#2. Observation \#3 also shows a transient FBO, which disappears for a few hundreds seconds during the observation.

\subsubsection{Ejection of lobes \#2}
Observations \#5 and \#6 started on MJD 51340.839 and 51340.909, respectively, shortly (85-95 min and 2-2.6 hr, respectively) after the ejection of NE and SW lobes \#2. These observations lasted 2.4 ks and 2.1 ks, respectively. 
The PDS corresponding to the above observations both show a weak flat-topped noise component and no LFQPOs, and two kHz QPOs at $\approx$670 and 950 Hz (not shown in the PDS in Fig. \ref{fig:PDS}). We note that kHz QPOs are frequently observed in both Sco X-1 and other Z-sources, and appear associated with different types of PDS shapes, not exclusively with flat-topped noise. In the HID shown in Fig. \ref{fig:CCD} the approximate location of the launch of the lobes \#2 is marked by a red-edged white diamond, and is found in correspondence with the HA, i.e. the HB-to-NB transition. The data from observation \#6 occupy approximately the same region of the HID. We note that while HBOs are commonly observed both on the HB and in correspondence with the HA in other Z-sources (see, e.g., \citealt{Homan2002} on GX 17+2), the lack of HBOs in correspondence with the HA in the data considered here is remarkable. Even though it is always possible that a QPO is present but not detected because too weak, a QPO non-detection in our very high S/N ratio data suggests that the HBO more likely disappears when Sco X-1 transitions from HB to NB.

\subsubsection{Launch of URF \#1}
Observations \#10 and \#11 started on MJD 51341.291 and MJD 51341.358, respectively, shortly before and after the first ejection of the URF, which occurred at $\approx$51341.32. Such observations lasted approximately 900s and 9ks, respectively. Observation \#10 ended shortly before the ejection of the URF, and the corresponding PDS shows a faint flat top noise and no sign of LFQPOs (but hints of twins KHz QPOs). Observation \#11, instead, started 30-80 minutes after the first ejection of the URF, and shows a strong NBO simultaneous to an HBO (at $\approx$5.8 and $\approx$44.7 Hz, respectively).  Such an NBO seems to persists over the two satellites snapshots\footnote{A snapshot is an unbroken pointing in a given direction, while an observation is formed by either one single snapshot, or by a collection of consecutive snapshots of a single target that does not generally span more than a few hours. In the case of RXTE, a single snapshot exposure roughly corresponds to the visibility window of the target along a given orbit, i.e. 1-2ks in the case of Sco X-1.} of observation \#11, with a steady centroid frequency of $\approx$5.8 Hz (see panels 11 in Fig. \ref{fig:PDS}). 
In the HID shown in Fig. \ref{fig:CCD} the approximate location of the launch of URF \#1 is marked by a red-edged blue star, and is found in correspondence with the centre of the NB.

We see that the PDS extracted from the subsequent RXTE observation (observation \#12), shows a transient FBO  detected for about 400s during the 1.9 ks of exposure. The PDS from observation \#13 shows an NBO at $\approx$9 Hz that erratically evolves into an FBO (with centroid frequency $\approx$ 12Hz)\footnote{This is a known characteristic of NBOs in Sco X-1, and such a behaviour has been reported by, e.g., \citealt{Casella2006}.}. Observation \#14 shows a transient NBO at $\approx$8 Hz, which progressively becomes fainter and disappears by the end of the observation. 

\subsubsection{Launch of URF \#2}
Observation \#15 started shortly before the second ejection of the URF and lasted 11.5 ks, and covers the URF ejection time. The PDS extracted from this observation shows a very strong NBO simultaneous with an HBO, that persist for the three snapshots of observation \#15 with a stable centroid frequency of $\approx$6 Hz. The PDS from the subsequent RXTE observation (observation \#16, taken $\sim$4 hours after the ejection of the URF), shows instead a faint flat-topped noise component, with no LFQPOs. In the HID shown in Fig. \ref{fig:CCD} the exact location of the launch of the URF \#2 (corresponding to observation \#15) is marked with a white stars, which is found along the NB, towards its soft end.
Observation \#17 shows a transient FBO with a centroid frequency of $\approx$ 17 Hz, which evolves into an NBO at the end of the observation. The appearance of simultaneous NBOs and HBOs have been reported already in the past (for sources observed in the RXTE era, see e.g. \citealt{vanderKlis1996}, \citealt{vanderKlis1997} and \citealt{Yu2007a} for Sco X-1; \citealt{Jonker2000} for GX 340+0; \citealt{Wijnands1996} and \citealt{Homan2002} for GX 17+2; \citealt{Jonker2002} and \citealt{Sriram2011} for GX 5-1; \citealt{Piraino2002} for Cyg X-2), but information about where exactly in the HID this QPO configuration appears is limited. \cite{vanderKlis1996} reports an NBO+HBO along the NB from data collected in 1996. These QPOs were detected all along the NB, but were most prominent half way through it. \cite{Homan2002} presented the timing of the Sco-like source GX 17+2 and showed that the NBO+HBO pair appeared in the soft end of the NB, shortly before the SA, i.e. the NB-to-FB transition. Our finding are  consistent with these results.

\smallskip
\noindent
From the above analyses we conclude that there is good evidence for an association between X-ray power spectra which combine NBO and HBO and the launch of the URF, and for an association between the hard apex and the corresponding flat-topped noise PDS and the launch of the lobes. We note, however, that the association between the launch of the URFs and the appearance of a NBO+HBO in the X-ray PDS is based on one single event for which both X-ray and radio data are available. Therefore, we cannot completely exclude this associations being  coincidental. 

\begin{table*}	
\centering									
\caption{Summary of the radio and X-ray timing behaviour during the radio observations reported F2001a,b. Note that a few of the RXTE observations considered (marked with a * in the \textit{Duration} column) consist in more than one snapshot, therefore the actual exposure is shorter than the  time interval covered by the pointing.}
\setlength\tabcolsep{3pt}
\begin{tabular}{p{2.4cm} p{4.8cm} p{5.5cm} p{0.8cm} p{2.2cm} p{0.8cm}}

\hline \hline			
Time					 &	Radio event											&	X-ray timing event 											&	Obs. \#		& RXTE Obs. ID	& Duration\\
(MJD)					 &	(from F2001a,b)							&	(\textit{RXTE} data)   		&			&		& [ks]\\
\hline		\hline

51339.254$\pm$0.003                & 	{\bf Ejection of lobes \#1}				&																&		&   	&		\\

\arrayrulecolor{gray}\hline
51339.384                & 														& Strong NBO + HBO  	&	1	& 40706-02-01-00  	&	1.5	\\
               & 														& (< 3 hr later than ejection of lobes \#1)		&		&   	&		\\

\hline
51339.645                & 														& Blurred NBO 														&	2	&  40706-02-03-00 	&	1.8	\\

\hline
51339.977                & 														& Blurred NBO + FBO										&	3		&  40706-02-06-00 	&	3.0	\\
\hline
51340.39$\pm$0.02         & Core flare C1										&																&			&		\\

\hline
51340.425			     & 														& Very blurred NBO 			&	4		& 40706-01-01-00  	&	3.5	\\
			     & 														& (< 20-80 min later than core radio flare C1)			&			&   	&		\\

\hline
51340.779$\pm$0.004    			 & {\bf Ejection of lobes \#2} 					&																&			&   	&		\\

\hline

51340.839				 &														& flat-topped noise												&	5		&  40706-02-09-00 	&	2.4	\\
                & 														& (< 85-95 min later than ejection of lobes \#2)												&			&   	&		\\


\hline
51340.86$\pm$0.02 			 & Core flare C2										&																&			&   	&		\\

\hline
51340.909                & 														& flat-topped noise												&	6		& 40706-02-08-00  	&	2.1	\\

\hline
51340.975			 	 &														& flat-topped noise												&	7		&  40706-02-10-00 	&	11.5*	\\

\hline
51341.159				 &														& flat-topped noise												&	8		&  40706-02-12-00 	&	1.5	\\

\hline
51341.225                & 														& flat-topped noise												&	9		&  40706-02-13-00 	&	1.6	\\

\hline
51341.291				 & 														& flat-topped noise												&	10		&  40706-02-14-00 	&	0.9	\\

\hline
51341.32 $\pm$0.02   & Core flare C3										&																&			&   	&		\\
                		 & Correlated with lobe flares N3 and S3 	&																&			&   	&		\\
						 & $\rightarrow$ \textbf{Ejection of URF \#1}		& 																&			&   	&		\\
\hline
51341.358                & 														&	Strong NBO + HBO 	&	11		&  40706-01-02-00 	&	9.0*	\\
                & 														&	(< 30-80 min later than Core flare No. 3)		&		&   	&		\\
\hline
51341.42 $\pm$0.02   & Lobe flare N3										&																&			&   	&		\\
                         & NE lobe \#2 slows down (?)								&																&			&   	&		\\
\hline
51341.577                & 														&	Faint FBO													&	12		&  40706-02-16-00 	&	1.9	\\

\hline
51341.699                & 														&	NBO	+ hint of FBO									&	13		&  40706-02-17-00 	&	1.8	\\

\hline
51341.906                & 														&	Very blurred NBO													&	14		&  40706-02-18-00 	&	2.2	\\
\hline
51341.973                & 														& Strong NBO + HBO 		&	15		&  40706-02-19-00 	&	11.5*	\\
                & 														& \textbf{(Simultaneous with ejection of invisible flow \#2)}		&		&   	&		\\

\hline
51341.99$\pm$0.02  & Core flare C4										&																&			&   	&		\\

                		 & Correlated with lobe flare N4 			&																&			&   	&		\\
						 & $\rightarrow$ \textbf{Ejection of the URF \#2}		& 																&			&   	&		\\

\hline
51342.13$\pm$0.02  & Lobe flare S3										&																&			&   	&		\\

\hline
51342.158				 &														& flat-topped noise												&	16		&  40706-02-21-00 	&	1.6	\\	

\hline
51342.291                & 	Low Core radio flux									& NBO + FBO 												&	17		&  40706-02-23-00 	&	1.0	\\
\hline
51342.3$\pm$0.02  & Lobe flare N4										&																&			&   	&		\\                

\arrayrulecolor{black}\hline 
\end{tabular}\label{tab:log}
\end{table*}																	

\section{Discussion}\label{sec:discussion}



Using techniques and knowledge developed since the observations were performed in the late 1990s, we have been able to re-analyse the X-ray data of Sco X-1 in order to try and look for connections between the two apparently different forms of relativistic ejection associated with the system. We turn first to the ultra-relativistic flows, which we remind the reader are not directly observed, but whose presence is inferred from apparently causally connected flaring in the core and subsequently in the approaching and receding lobes.
There is a strong hint from our analyses that these events occur when both the NBO and HBO are present in the X-ray power spectrum. 
Specifically, the first ejection of the URF occurred in between Obs. \#10 and Obs. \#11, the PDS of which show flat-topped noise, and a strong NBO simultaneous with an HBO (that persisted for at least the 9ks of the observation), respectively. During this second observation Sco X-1 was found in the NB. One RXTE observation (Obs. \#15) occurred  at precisely the time of radio flare C4, which we associate with the launch of the second URF in our data set. The X-ray PDS shows a strong NBO simultaneous with a high-frequency HBO ($\approx$6.42 and $\approx$43Hz, respectively). Both the URF ejection and RXTE observation \#15 occur along the NB, closer to its softer end (Fig. \ref{fig:CCD}). Unfortunately, while the URF ejection happens during the RXTE observation, the large time uncertainties on the time of the event, as well as the limited exposure of the RXTE observations, prevented us to determine whether the ejection of the URF preceded or followed the appearance of the NBO in the PDS. In this regard, we note that in general the uncertainties on the radio events times, i.e. 0.02 days, is comparable to the average lengths of an RXTE snapshot of Sco X-1. At the approximate time when the URFs were launched, Sco X-1 was overall moving through the HID from the the HB to the NB. However, a closer inspection of the HID (details not shown) reveals that multiple back and forth transitions occurred close to the launch of URF \#2 during observation \#15, thus possibly during the launch of URF \#1 as well. 

How common is this particular state? We have inspected a large number of \textit{RXTE} archival observations of Sco X-1\footnote{We considered only observations performed in standard modes, which could be used to produce PDS covering frequencies below 64 Hz. For instance, we did not use \textsc{GoodXenon} mode data taken when the source count rate exceeded  8000 cnt/s (Motta et al. 2017, submitted).} and found that while HBOs simultaneous with NBOs seem to be relatively short lived (with the HBO frequently disappearing in a few ks, leaving an isolated NBO), isolated NBOs seem to have a longer life-time, with a tendency of moving at slightly higher frequencies with time. In 350 archival observations, $\approx$45\% showed an NBO. Of these, $\approx 20\%$ appeared simultaneously with an HBO. While some NBOs ($< 10\%$) were significantly detected for only a few minutes, the majority were detected for the entire duration of one or more consecutive RXTE snapshots, and often ($\approx 60\%$ of the cases) over 1 or more observations (assuming that the NBO did not disappear in between snapshots/observations). This suggests that the NBO average life-time is approximately 0.5 day. In only $\approx 4\%$ of the cases the simultaneous NBO+HBO were detected over more than one RXTE observations, while in the remaining cases the NBO+HBO were visible for one RXTE observation or less (typically around 2ks). This suggests that while some NBOs+HBOs could survive for up to $\approx$0.5 day (similarly to most NBOs), the majority of them likely remained visible for less than 3 hours. This implies that the simultaneous NBO+HBO observed shortly after the first ejection of the URF (RXTE Obs. \#11, PDS in panel 11 in Fig. \ref{tab:PDS_fits}) could be effectively related to the ejection event. 

It has been previously established that any time an NBO+HBO PDS configuration appears, the HBO is detected at a frequency very close or coincident with the maximum frequency reached by HBOs in Sco X-1 (see, among others, \citealt{vanderKlis1996}, \citealt{vanderKlis1997} and \citealt{Yu2007a} for Sco X-1; \citealt{Jonker2000} for GX 340+0; \citealt{Wijnands1996} and \citealt{Homan2002} for GX 17+2; \citealt{Jonker2002} and \citealt{Sriram2011} for GX 5-1; \citealt{Piraino2002} for Cyg X-2). 
As long as the QPO production mechanism maintains a dependence on the radius where the QPO is produced - which is the case for all the QPO models proposed so far (see e.g. \citealt{vanderKlis2005}) - the presence of a maximum QPO frequency (or of a ``saturation'' frequency) necessarily corresponds to the minimum radius at which a QPO can be produced. This minimum must be linked to either the Alfv\'en radius, or the surface of the NS, or the innermost stable circular orbit, depending on the NS equation of state and on the NS magnetic field. This implies that the ejection of the URF might occur when the accretion flow surrounding the NS reaches the minimum distance from the compact object. 

\begin{figure}
\includegraphics[width=0.45\textwidth]{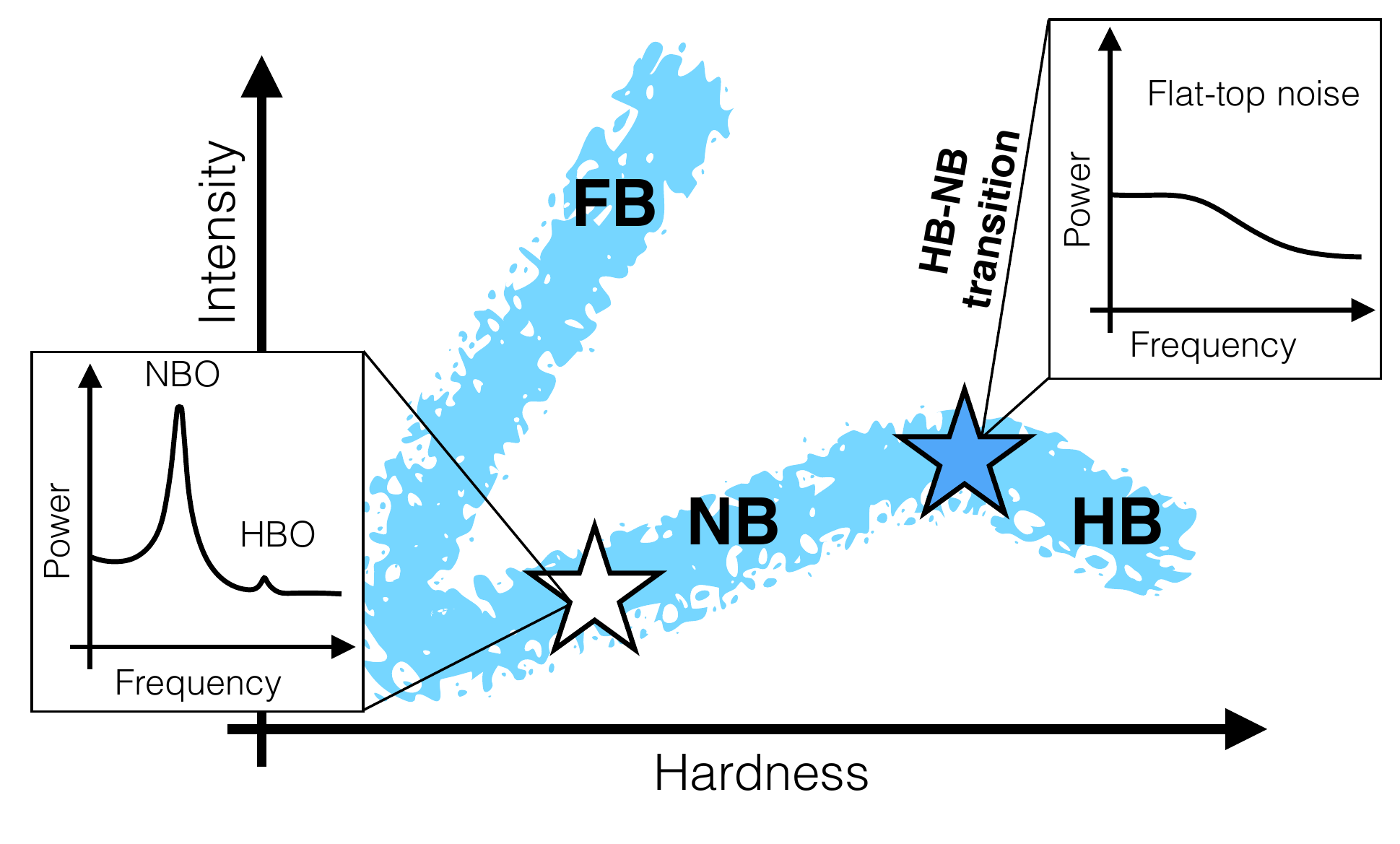}
\caption{Sketch of the disc-jet coupling in Sco X-1. The PDS shapes corresponding to the HB-NB transition (hard apex) and to the NB at the time of the URF ejections are shown in the insets. The location of the launch of ballistic ejections and of the URFs on the HID are indicated with a blue and white star, respectively.}
\label{fig:ske}
\end{figure}

Regarding the slower-moving, radio-emitting lobes, the results are less clear. There is no RXTE coverage at the time of the ejection of lobes \#1. Lobes \#2 may be associated with a PDS characterised by flat-topped noise, but the uncertainty in whether or not the lobes were moving ballistically or decelerated adds considerable ambiguity. Using Occam's razor and assuming that the motions were ballistic, the lobes were ejected during a phase when the power spectrum was evolving from a blurred NBO to flat-topped noise, but considerably closer ($\approx$1.5 hr as opposed to $\approx$8.5 hr) to a flat-topped noise PDS `phase', which corresponds to the hard apex in the Z-track. Interestingly, \cite{Church2014} suggested that the launch of a jet is expected at this stage of the evolution along the Z-track due to an increase of the mass accretion rate in the direction of the HA, which causes high radiation pressure and thus a vertical deflection of the accretion flow, which is launched in a jet. In this context the lack of HBOs at the time of the ballistic ejections might support the idea that the inner disc might be disrupted in this phase, which would stop relativistic precession, or whatever mechanism is responsible for the generation of the HBOs. The occurrence of the different type of jet ejections in Sco X-1 in relation to states and transitions identified along the Z-track is summarized in schematic form in Fig. \ref{fig:ske}. 

The above results leads on to the question of the general nature of the core flaring. 

\subsection{On the nature of the core flaring}

Our findings imply that each ejection event covered by the radio monitoring corresponds to a core flare. This is obviously true for the URFs - the presence of which is inferred based on the correlated core/lobes flares - but also for at least the lobe ejection covered by the radio monitoring (the lobe ejection \#2). The broad paradigm used for most jet sources is based on such an interpretation -- core radio flares indicate ejections.
However, we note that not all the core flares in Sco X-1 are obviously associated with an ejection event. For instance, core flare C1 is apparently not associated with an ejection of either type. Shortly after core flare C1, though, the radio lobes were no longer detected, therefore it is possible that flare C1 corresponded to the ejection of an invisible flow that did not cause a detectable flare in the lobes. Of course, the data do not allow us to exclude that flare C1 is linked to an event of some sort that does not have a counterpart in radio or in the X-rays.

From our findings and based on what has been reported by F2001a,b, it is clear that intrinsically different ejection events (lobe ejections versus ejections of URFs) are linked to core flares that do not differ from each other in any obvious way, even when radio spectral information is available. For instance, results by \cite{Fomalont2001} show that the radio spectral index of the radio core measured during a lobe ejection does not significantly differ from the radio spectral index measured during an inferred URF launch. Similarly, the evolution of the core spectral index after an ejection of either kind does not show any obvious difference in the two cases. Furthermore, the radio spectra extracted from the radio core and lobes emission also show very similar spectral index values, scattered over broadly overlapping intervals. Additionally, as already noted by F2001a,b, the flares associated to the lobes - and especially those from the NE lobe (which is launched towards the observer) - are often as intense or even more intense than those from the core. This means that for typical lower angular resolution radio observations (i.e. non-VLBI), it is impossible to tell with certainty whether radio flaring is associated with the core or, delayed, associated with the lobes. 

Drawing a comparison with black hole binaries, it is known that relativistic ejection events are associated (although perhaps not solely) with transitions between accretion states (Fender et al. 2004). It has further been suggested that the most precise indicator of the moment of relativistic jet launch in BH may be the presence of `type-B' QPOs (the BH equivalent of the neutron star NBO) in the power density spectra, although establishing a direct connection has been difficult (\citealt{Fender2009}, \citealt{Miller-Jones2012}). There are clear analogies here with the particular power spectral state, indicated by the simultaneous presence of the NBO and HBO, occurring along the NB (see Fig 5). However, nothing like the URFs discussed here has ever been observed in a confirmed black hole systems, which leads to the question: do they exist in BH or are they unique to NS? If the latter, this implies that they are probably connected to the existence of a solid surface and/or a surface magnetic field. If the former, there may be evidence for them in existing data sets, although the majority of radio data for black holes is unresolved flux monitoring which will remain ambiguous. Furthermore, a Type-B + type-C QPO configuration (i.e. the equivalent of an NBO+HBO configuration) is only rarely observed in BHs systems, and seems to be related to high accretion rates (i.e. close to the Eddington rate, see, e.g., \citealt{Motta2012} and \citealt{Motta2014a}). Finally, we note that while the URF needs to have a high speed, it does not necessarily need to carry that much more energy than the slower-moving lobes, since the only constraint we can place on the delivered energy is that associated with the lobes re-brightening. Straightforward minimum-energy arguments can only constrain that this be $\geq 10^{35}$ erg s$^{-1}$, comparable to the value estimated for the apparently related phenomena in Cir X-1 \citep{Fender2004b}.

In conclusion, we have found the first evidence, albeit tentative,  for a connection between a particular accretion state and the launch of ultra-relativistic outflows in a neutron stars X-ray binary. The physical interpretation of these unseen flows remains uncertain, as does the question of whether such a phenomenon occurs in black holes and how important it is energetically for the processes of accretion and ejection around relativistic objects. Future high-resolution observations of other neutron star and black hole X-ray binaries, as well as study of archival data on AGN, should shed further light on this question.

\bigskip
\section{Acknowledgements}

SEM and RPF acknowledge the anonymous referee, whose comments help improving this work. SEM acknowledges the Violette and Samuel Glasstone Research Fellowship programme and the UK Science and Technology Facilities Council (STFC) for financial support.

\appendix

\section{RXTE data power density spectra}\label{App:PDS}



\begin{figure*}
\centering
\includegraphics[width=0.95\textwidth]{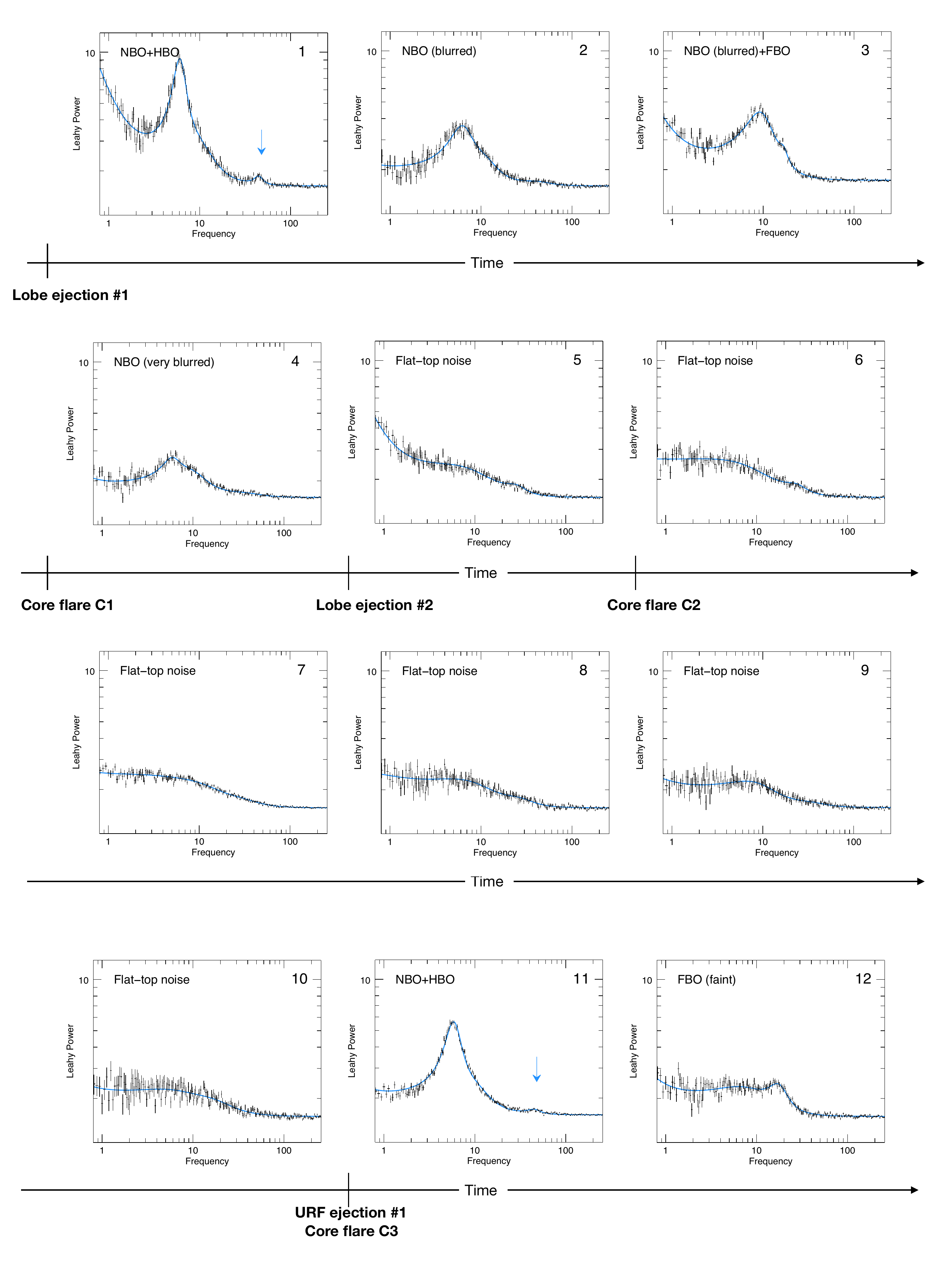}
\caption{PDS extracted form the \textit{RXTE} data taken between MJD 51339 and MJD 51343. For each panel we show the Leahy normalized PDS. The numbers reported at the top-right corner of each panel correspond to the numbers reported in Fig. \ref{fig:fomalont} (bottom panel) and in Tab. \ref{tab:log}. The blue arrow, when present, indicate the position of the HBO (which always shows an amplitude significantly lower than that of the NBO).}\label{fig:PDS}
\end{figure*}

\renewcommand{\thefigure}{\thesection.\arabic{figure} (Continued)}
\addtocounter{figure}{-1}

\begin{figure*}
\centering
\includegraphics[width=0.95\textwidth]{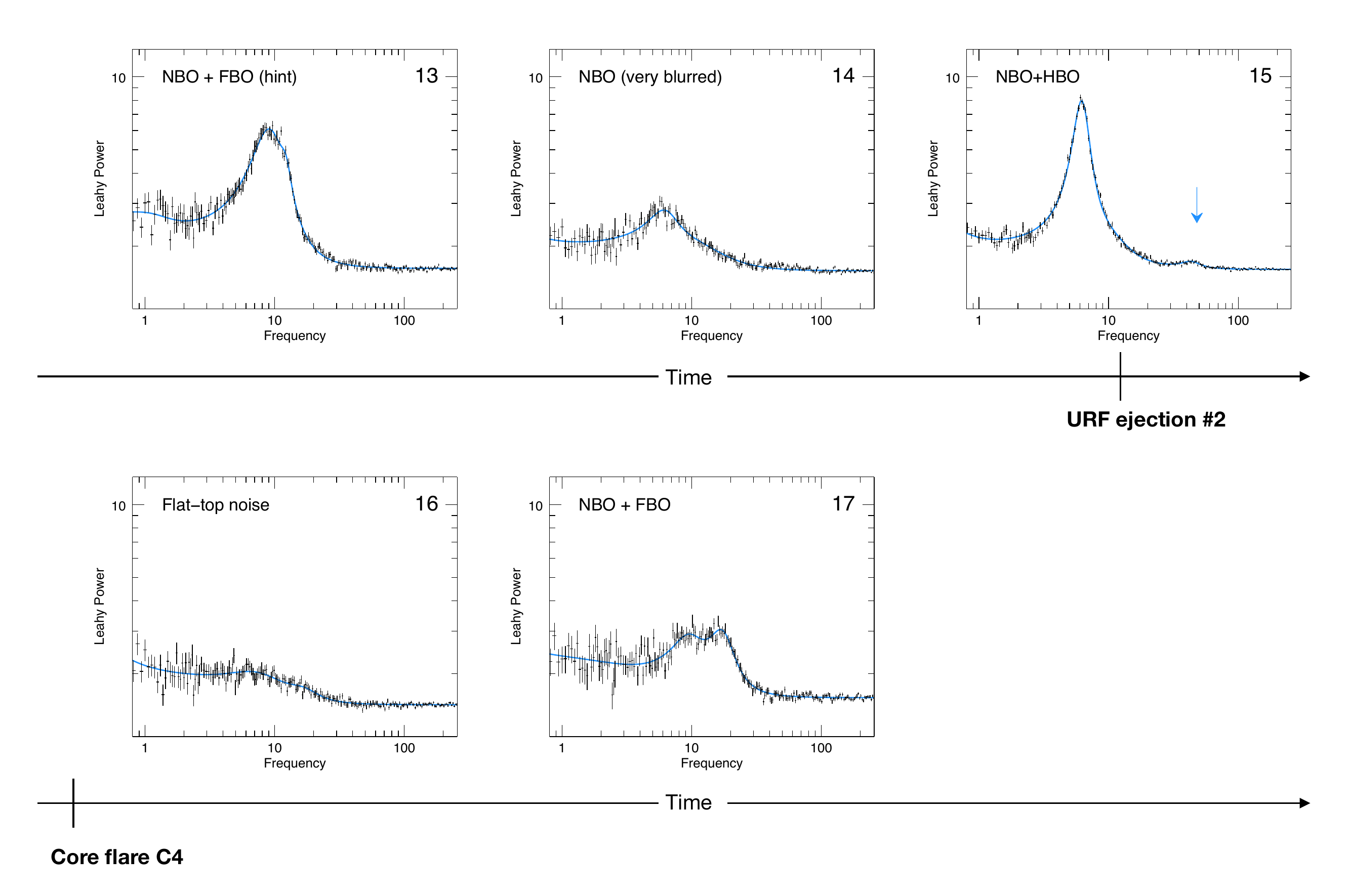}
\caption{ }
\end{figure*}




\bibliographystyle{mnras.bst}
\bibliography{biblio}

%





\label{lastpage}
\end{document}